\documentclass[12pt,aps,amsmath,latexsym,amsfonts]{JHEP3}

\usepackage{epsfig}
\usepackage{amsfonts,amssymb,amsmath}
\usepackage[hang]{subfigure}

\newcommand{\be}{\begin{equation}}
\newcommand{\ee}{\end{equation}}
\newcommand{\bea}{\begin{eqnarray}}
\newcommand{\eea}{\end{eqnarray}}

\def\half{\textstyle{\frac{1}{2}}}

\def\quarter{\textstyle{\frac{1}{4}}}

\title{Gauss-Bonnet Holographic Superconductors}
\author{
Luke Barclay\thanks{Email: luke.barclay@durham.ac.uk} , 
Ruth Gregory\thanks{Email: r.a.w.gregory@durham.ac.uk} , 
Sugumi Kanno\thanks{Email: sugumi.kanno@durham.ac.uk} ,
Paul Sutcliffe\thanks{Email: p.m.sutcliffe@durham.ac.uk}
\\
{\it Centre for Particle Theory, Department of Mathematical Sciences\\
Durham University, South Road, Durham, DH1 3LE, UK}}

\abstract{
We study holographic superconductors in five dimensional
Einstein-Gauss-Bonnet gravity both numerically and analytically.
We find the critical temperature of the superconductor decreases
as backreaction is increased, although the effect of the 
Gauss-Bonnet coupling is more subtle: the critical temperature
first decreases then increases as the coupling tends towards the 
Chern-Simons value in a backreaction dependent fashion.
We compute the conductivity of the system, finding the
energy gap, and show that the effect of both backreaction
and higher curvature is to increase the gap ratio $\omega_g/T_c$,
thus there is no universal relation for these superconductors.
}

\keywords{ads/cft, holography}
\preprint{DCPT-10/41}

\begin{document}
\newcommand{\zed}{$\mathbb{Z}_2$}

\section{Introduction}

The gauge gravity correspondence, \cite{Malda}, provides a fascinating tool 
to explore strongly coupled field theory. Recently, it has been
applied to condensed matter systems yielding 
interesting qualitative results for systems exhibiting a
superconducting phase \cite{HHrev}. In these, the bulk ``classical'' 
theory has gauge and charged scalar (or fermi) fields,
and a black hole provides a finite temperature.
Typically, no hair theorems, \cite{nohair}, would lead us to 
believe that the scalar field must be in its vacuum even in
a highly charged black hole background, however, such theorems
do not take account of the fact that in anti de Sitter spacetime
scalar fields can have an apparently tachyonic mass, provided it
is not too large, \cite{BF}; the confining properties of anti-de
Sitter (adS) spacetime in essence prevent the instability from 
setting in unless the wavelength
is sufficiently small. Once the scalar has a negative mode, the
usual conditions of no-hair theorems cease to hold, and it becomes
possible for the scalar to condense out of the vacuum in an
analogous fashion to the condensation of SU(2) t'Hooft Polyakov 
fields outside a magnetic Reissner-Nordstrom solution \cite{MAGRN}.

The basic picture is that at a sufficiently low temperature, it
becomes energetically feasible for a charged scalar to acquire an
expectation value near the event horizon of the black hole, \cite{Gub}, 
spontaneously breaking the gauge symmetry and screening the charge 
and mass of the black hole. 
Because the asymptotic true vacuum is symmetric, the scalar
has a power law fall-off near the boundary, and the coefficient
of this fall-off can be interpreted as a condensate in the 
boundary theory. There is a reasonably complete understanding of
these systems, including varying scalar mass and potential, the
gauge group, the number of spacetime dimensions, as well as
having magnetic fields present and the stability of the 
system \cite{Gub,HHH,HorRob,zeroT,color,DIM,magnetic,KS}.
While most of these models are ``bottom up'', in the sense of
being empirically constructed, similar models have been found
from the top down perspective \cite{STSC}, and
it is therefore of interest to consider more general stringy
aspects of these models.

In a previous paper, some of us, \cite{GKS}, explored 
the stability of these empirical models to leading order
corrections, by including on the gravitational side
higher curvature terms, specifically, the Gauss-Bonnet (GB)
invariant \cite{GBL} which is believed to be the ${\cal O}(\alpha')$
correction to low energy string gravity \cite{LESG}. 
We found that many of the qualitative 
features of the holographic superconductors were stable under 
higher curvature corrections, however, upon studying the conductivity,
a more interesting story emerged. In \cite{HorRob}, Horowitz and Roberts 
studied $2$ and $3+1$-dimensional holographic superconductors for a 
variety of bulk scalar masses. They discovered that the energy gap
typically present in the real part of the conductivity, $\omega_g$,
was always about eight times as large as the critical temperature
of the superconductor:
\be
\omega_g \simeq 8 T_c
\ee
this suggested that in spite of the different dimensions of the
boundary condensate, there was some universal mechanism 
governing their superconductivity. Upon adding in higher
curvature corrections however, this ``universal'' relation 
completely disappeared. This analysis was confirmed for more 
general models in \cite{GBscs}, and although the expressions for
conductivity used in these papers are not strictly accurate (see
comments in section \ref{sec:cond} and appendix \ref{AppA})
the gap result is. The analyses presented 
in \cite{GKS,GBscs} are however in the probe limit, i.e.\
where the matter fields do not backreact gravitationally
on the spacetime, and since backreaction does typically alter
the critical temperature, it is likely that a full computation
will alter the relation between $\omega_g$ and $T_c$, potentially
restoring the universal relation, at least for some value of the
gravitational coupling for each $\alpha$. In \cite{Betti},
the stability of the scalar condensate to higher curvature corrections
was tested, and it was found that backreaction lowered $T_c$, as with
Einstein gravity, thus making condensation harder. There was no 
qualitative difference however with the probe results.

In this paper we include the full effect of
gravitational backreaction on the holographic superconductor
with higher curvature terms. We find analytic bounds on the critical
temperature, and cross-check these against exact results obtained
by numerical computation. We then numerically compute the conductivity, 
and demonstrate that the conductivity does not in fact have a universal
gap. Indeed, we find that even in Einstein gravity
there is no universal gap once backreaction is taken into account.
The organization of the paper is as follows:
We first review the GB holographic superconductor in section \ref{sec:SC},
discussing the bulk model, clarifying possible ambiguities in the choice
of mass of the scalar field and setting out some of the basic properties
of solutions to the system of equations.
In section \ref{sec:BR} we analyse the bulk equations analytically,
providing bounds on the critical temperature and present numerical
results for the condensation and critical temperature of 
the holographic superconductor. We then investigate the conductivity
in section \ref{sec:cond}, first deriving the backreacted conductivity
equation for GB gravity, then solving it numerically. Finally, we conclude
in section \ref{sec:disc}.

\section{The Gauss-Bonnet bulk superconductor}
\label{sec:SC}

We begin with the Einstein-Gauss-Bonnet (EGB) gravitational action coupled
to a massive charged complex scalar field and a U(1) gauge field:
\begin{eqnarray}
S&=&\frac{1}{2\kappa^2}\int d^5x \sqrt{-g} \left[
-R + \frac{12}{L^2} + \frac{\alpha}{2} \left(
R^{abcd}R_{abcd} -4R^{ab}R_{ab} + R^2
\right) \right]\nonumber\\
&&\hspace{3.5mm}+\int d^5x\sqrt{-g}\left[
-\frac{1}{4}F^{ab}F_{ab}+|\nabla_a\psi -iqA_a\psi|^2
-m^2|\psi|^2
\right]\,
\label{action}
\end{eqnarray}
where $g$ is the determinant of the metric, and
$R_{abcd}$, $R_{ab}$ and $R$ are the Riemann curvature tensor,
Ricci tensor, and the Ricci scalar, respectively.
We take the Gauss-Bonnet coupling constant $\alpha$ to be positive, and
the negative cosmological constant term, $-6/L^2$, has been written
in terms of a length scale, $L$. Note $\kappa^2 = 8\pi G_5$ gives
an explicit Planck scale, and $q$ is the charge, and $m$ the mass,
of the scalar field $\psi$\footnote{Note that we follow Horowitz et
al., \cite{HHH}, in taking a purely quadratic potential for the scalar
field.}.

The equations of motion can be readily derived as:
\be
\label{GBeqns}
R_{ab} - {\half} R g_{ab} + \frac{6}{L^2} g_{ab}
-\alpha \left [ H_{ab} - \quarter H g_{ab} \right ] = 8\pi G T_{ab}
\ee
where
\be
\label{Hdef}
H_{ab} = R_a^{\;cde} R_{bcde} - 2 R_{ac} R^c_b - 2 R_{acbd} R^{cd}
+R R_{ab} \;,
\ee
$T_{ab}$ is the matter energy momentum tensor 
\be
\label{tabdef}
T_{ab} = 2 {\cal D}_{(a} \psi^\dagger {\cal D}_{b)} \psi
-F_{ac} F_b^{\;c} - \left [
|{\cal D}_c \psi |^2 - \quarter F_{cd}^2 - m^2 |\psi|^2
\right ]g_{ab} \;,
\ee
and ${\cal D}_a = \nabla_a - iqA_a$ is the gauge covariant derivative.

In the absence of any matter sources, the EGB equations have a pure
adS solution
\begin{eqnarray}
ds^2 = \frac{r^2}{L_e^2}\left [ dt^2 - (dx^2+dy^2+dz^2) \right ]
- \frac{L_e^2}{r^2} dr^2
\end{eqnarray}
with
\begin{eqnarray}\label{Leff}
L^2_{\rm e}=\frac{2\alpha}{1-\sqrt{1-\frac{4\alpha}{L^2}}}
\to  \left\{
\begin{array}{rl}
L^2   \ , &  \quad {\rm for} \ \alpha \rightarrow 0 \\
\frac{L^2}{2}  \ , &  \quad {\rm for} \  \alpha \rightarrow \frac{L^2}{4}
\end{array}\right.
\,.
\end{eqnarray}
Thus the actual curvature of the adS spacetime 
is renormalized away from the cosmological constant scale, $L$, 
once $\alpha$ is nonzero. Since $L_e<L$, one could interpret 
switching $\alpha$ on as strengthening gravity, in the sense that
a shorter lengthscale corresponds to stronger curvature.

To examine holographic superconductivity, we look
for plane-symmetric black hole solutions with or without a nontrivial
scalar, but with a nonzero charge. Taking the metric ansatz
\begin{eqnarray}
ds^2 = f(r)e^{2\nu(r)}dt^2 - \frac{dr^2}{f(r)} 
- \frac{r^2}{L_e^2}(dx^2+dy^2+dz^2)
\end{eqnarray}
in the absence of a scalar field there is an analytic charged
black hole solution, \cite{GBBH}, with $\nu=0$, and
\bea
\label{phibk}
A &=& \phi(r) dt 
= \frac{Q}{r_+^2} \left ( 1 - \frac{r_+^2}{r^2} \right) dt \\
f(r) &=& \frac{r^2}{2\alpha} \left [ 1 - \sqrt{
1 - \frac{4\alpha}{L^2} \left ( 1 - \frac{r_+^4}{r^4} \right )
+ \frac{8\alpha\kappa^2 Q^2}{3r^4 r_+^2} \left ( 1 - \frac{r_+^2}{r^2}
\right )} \right]
\label{fbk}
\eea
where $Q$ is the charge of the black hole (up to a geometrical factor
of $4\pi$), and $r_+$ is the event horizon, which determines the ``ADM''
mass of the black hole \cite{ADMass}. In order to avoid a naked
singularity, we need to restrict the parameter range as $\alpha\leq L^2/4$.
Note that in the Einstein limit ($\alpha\rightarrow 0$), the solution 
(\ref{fbk}) goes to the Reissner-Nordstr\"{o}m adS black hole: 
\begin{eqnarray}
f(r)=\frac{r_+^2}{L^2}\left(\frac{r^2}{r_+^2}-\frac{r_+^2}{r^2}\right)
+\frac{2\kappa^2 Q^2}{3r_+^4}\left(\frac{r_+^4}{r^4}-\frac{r_+^2}{r^2}\right)
\label{r:rn}\,.
\end{eqnarray}
In the Chern-Simons limit, $\alpha = L^2/4$, (so called because
in odd dimensions the gravitational lagrangian becomes the potential 
for the Euler density in one dimension up, see e.g.\ \cite{BTZn})
the Newtonian potential takes the simpler form:
\be
\label{CSBH}
f(r) = \frac{2r^2}{L^2} - \frac{2r_+^2}{L^2}
\sqrt{1+2 L^2 \frac{\kappa^2 Q^2}{3r_+^6}
\left ( 1 - \frac{r_+^2}{r^2}\right)} \,.
\ee

The superconducting phase corresponds to a ``hairy'' black hole, where 
the scalar field has condensed out of its symmetric state and screens
the charge of the black hole.  We can see that this will happen
at sufficiently low temperature from Gubser's rough argument, \cite{Gub},
using the scalar `effective mass' $m_{\rm eff}^2=m^2-q^2\phi(r)^2/f(r)$.
For low temperature black holes, $f(r)$ increases slowly away from the
horizon, giving a large negative mass squared over a sufficient
range for an instability to set in. This is of course confirmed by
the numerical results.

For the purposes of our investigation, we wish to have a fixed mass 
in order to focus on the effects of backreaction and the higher
curvature terms. In \cite{HHH}, the mass of the scalar was given
in terms of the adS lengthscale ($m^2 = -(D-2)/L^2$
in $D$-dimensions), and thus fixed with respect to the cosmological
constant via the Einstein equations. In \cite{GKS,Betti} this same
value of the mass was also chosen with respect to the {\it cosmological
constant} scale $L$. However, in EGB gravity, the adS lengthscale is
dependent on both $L$ and $\alpha$ via (\ref{Leff}), and 
fixing the mass with respect to $L$ and not $L_e$ 
means that the dimension of the operator in the dual boundary
theory varies with $\alpha$. It is possible therefore that some of the 
phenomena observed in \cite{GKS,Betti} are a consequence of this
varying dimension, rather than intrinsic to the system,
and we therefore fix the mass relative to the asymptotic
adS lengthscale, $m^2 = -3/L_e^2$, in order that the boundary operator
has fixed dimension $3$.

In order to look for the hairy black hole solution, we take the
standard static ansatz for the fields:
\begin{eqnarray}
A_a=\phi(r) \delta_a^0\,,\hspace{1cm}\psi =\psi(r)\;,
\end{eqnarray} 
where without loss of generality $\psi$ can be taken to be
real. 
The full system of gravity and gauge-scalar equations of motion
is then obtained as\footnote{Note, the $ij$ component of the 
EGB equations is not independent via a Bianchi identity.}:
\begin{eqnarray}
&&\phi^{\prime\prime}+\left( \frac{3}{r}-\nu^\prime
\right)\phi^\prime -2q^2\frac{\psi^2}{f}\phi=0\,, \label{phieq}\\
&&\psi^{\prime\prime}+\left( \frac{3}{r}+\nu^\prime+\frac{f^\prime}{f}
\right)\psi^\prime +\left(\frac{q^2\phi^2}{f^2e^{2\nu}}
-\frac{m^2}{f} \right)\psi=0\,, \label{psieq}\\
&&\left(1-\frac{2\alpha f}{r^2} \right)\nu^\prime
=\frac{2\kappa^2}{3}r\left(
\psi^{\prime 2}+\frac{q^2\phi^2\psi^2}{f^2e^{2\nu}}\right) \label{nueq}\\
&&\left(1-\frac{2\alpha f}{r^2} \right)f^\prime+\frac{2}{r}f
-\frac{4r}{L^2} =-\frac{2\kappa^2}{3}r\left[
\frac{\phi^{\prime2}}{2e^{2\nu}}+m^2\psi^2+
f\psi^{\prime2}+\frac{q^2\phi^2\psi^2}{fe^{2\nu}} \right] \label{feq}
\end{eqnarray}
where a prime denotes derivative with respect to $r$.
These equations have several scaling symmetries, similar to
those noted in \cite{HHH}, although as we have explicitly kept
the Planck scale, an additional symmetry
corresponding to a rescaling of energy is present.

\begin{enumerate}

\item
$r \to ar$, $t, x^i \to at, ax^i$,
$L \to aL$, $q \to q/a$, $\alpha \to a^2 \alpha$, $A \to a A$.

\item
$r \to br$, $t \to t/b$, $x^i \to x^i/b$,
$f \to b^2f$, $\phi \to b\phi$.

\item
$\phi \to c \phi$, $\psi \to c\psi$, $q \to q/c$, $\kappa^2 \to \kappa^2/c^2$.

\end{enumerate}
We use these rescalings to set $L=Q=q=1$ for
numerical convenience. Note that in \cite{HHH}, the Planck
scale was set to unity a priori, hence the probe limit
corresponded to $q\rightarrow\infty$, as can be seen from the third
rescaling. Also in contrast to \cite{HHH}, we choose to fix $Q=1$,
so that we are explicitly holding the charge parameter fixed in all
our computations.

Finally, the Hawking temperature is given by
\begin{eqnarray}
T = \frac{1}{4\pi} f' (r)e^{\nu(r)}\bigg|_{r=r_+}\ ,
\label{hawking}
\end{eqnarray}
this will be interpreted as the temperature of the conformal 
field theory (CFT).

In order to solve our equations we need to impose boundary conditions
at the horizon and the adS boundary.

\noindent $\bullet$ Horizon:

The position of the horizon, $r_+$, is defined by $f(r_+)=0$. 
Demanding regularity of the matter fields and metric at the horizon 
gives:
\begin{eqnarray}
\phi(r_+)=0,\hspace{1cm}
\psi^\prime(r_+)=\frac{m^2}{f^\prime(r_+)}\psi(r_+) \ .
\end{eqnarray}
Then equations (\ref{nueq}) and (\ref{feq}) give:
\begin{eqnarray}
&&
\nu^\prime(r_+)=\frac{2\kappa^2}{3}r_+\left(
\psi^\prime(r_+)^2
+\frac{\phi^\prime(r_+)^2\psi(r_+)^2}{f^\prime(r_+)^2e^{2\nu(r_+)}}
\right)\\
&&
f^\prime(r_+)=\frac{4}{L^2}r_+
-\frac{2\kappa^2}{3}r_+\left(
\frac{\phi^\prime(r_+)^2}{2e^{2\nu(r_+)}}
+m^2\psi(r_+)^2
\right)
\end{eqnarray}

\noindent $\bullet$ Boundary:

As we want the spacetime to asymptote to adS in standard coordinates, 
we look for a solution with 
\begin{eqnarray}
&&\nu \to 0 \;\;\;,\;\;\;\;
f(r) \sim \frac{r^2}{L_e^2}\;\;\; {\rm as} \;\; r \to \infty\,.
\end{eqnarray}
Asymptotically the solutions of $\phi$ and $\psi$ are then found to be:
\begin{eqnarray}
\phi(r) \sim P - \frac{Q}{r^2}\,,\hspace{1cm} 
\psi(r) \sim \frac{C_{-}}{r^{\Delta_-}}+\frac{C_{+}}{r^{\Delta_+}}\,,
\label{r:boundary}
\end{eqnarray}
where $\Delta_\pm=2\pm\sqrt{4+m^2L_e^2}$ for a general mass $m^2$. 
In order to have a normalizable
solution we take $C_-=0$; $P$ and $C_+$ are then fixed by consistency
with the near horizon solution.
According to the adS/CFT correspondence, we can interpret 
$ \langle {\cal O}_{\Delta_+} \rangle \equiv C_{+}$, 
where ${\cal O}_{\Delta_+}$ is the operator with the conformal
dimension $\Delta_+$ dual to the scalar field.
As already mentioned, we want the dimension of the boundary operator
to remain fixed as we vary $\alpha$ and so we take $m^2 = -3/L_e^2$.
Thus $\Delta_+ = 3$ for our choice of mass, and we compute the solutions
of (\ref{phieq}$-$\ref{feq}) numerically, reading off the $r^{-3}$ 
fall-off of the scalar field to obtain $\langle {\cal O}_3 \rangle$ 
for a range of different temperatures.

\section{Backreacting superconductors}
\label{sec:BR}

As shown in \cite{GKS}, the scalar field condenses out of its
vacuum near the horizon of a sufficiently small black hole. Although
the mass chosen for the scalar in \cite{GKS} is different to that
used here ($m^{\prime2} = -3/L^2 \geq -3/L_e^2$) we still expect
a similar qualitative behaviour. Thus, in computing the dependence
of $\langle {\cal O}_{3} \rangle$ with $T$, we expect to see 
the characteristic curve depicting 
the condensation of $\langle {\cal O}_3 \rangle$ 
from being trivially zero above some critical temperature $T_c$, 
to some nonzero value below this critical temperature. 
Before proceeding with a full numerical analysis,
it is useful to obtain some analytic understanding of the phase transition 
and the critical temperature.

In order to find analytic bounds for the critical temperature,
we look at the scalar field equation near $T_c$. For temperatures
just below $T_c$, the scalar is only marginally away from its vacuum,
hence the metric and gauge field will have the form
(\ref{phibk}, \ref{fbk}), up to corrections of order ${\cal O} (\psi^2)$, 
thus the scalar field satisfies a linear equation (\ref{psieq})
with $f$ and $\phi$ taking their background values.

A crude upper bound for $T_c$ can be found by considering the 
variable $X = r^{2}\psi$, which satisfies to leading order: 
\be
X'' + \left ( \frac{f_0'}{f_0} - \frac{1}{r} \right ) X' + \left (
\frac{q^2\phi^2}{f_0^2} + \frac{3}{L_e^2f_0} - \frac{2f_0'}{rf_0} 
\right ) X = 0 \; .
\label{Xeqn}
\ee
At the horizon,
\be
X'(r_+) = \frac{X(r_+)}{4\pi T_c} \left ( \frac{8}{L^2} - \frac{3}{L_e^2} 
- \frac{8\kappa^2 Q^2}{3r_+^6} \right ) 
\ee
which is positive for small $\kappa^2 Q^2$ (taking $X(r_+)>0$ without
loss of generality). Since $X\sim 1/r$ as $r\to \infty$, the solution
must have a maximum for some $r$, which requires that
\be
\left ( \frac{q^2\phi^2}{f_0^2} + \frac{3}{L_e^2f_0} 
- \frac{2f_0'}{rf_0} \right ) >0
\label{upperbdfn}
\ee
at this point.  An examination of when this condition is violated provides
an upper bound for $T_c$. This bound is only reliable for weakly
gravitating systems, and figure \ref{fig:anbound} shows the upper bound
for no backreaction, and $\kappa^2 = 0.05$. For larger values of 
$\kappa^2$, the method fails because the behaviour of the function in
(\ref{upperbdfn}) qualitatively changes.

In order to obtain a lower bound, consider instead $Y = rX$, 
then manipulating the equation of motion for $Y$ implies 
that {\it if} a solution exists, then the integral
\be
\int_{r_+}^\infty \frac{1}{r^3}  \left [\frac{\phi_0^2}{f_0} + \frac{3}{L_e^2} 
+ \frac{3f_0}{r^2} - \frac{3f_0'}{r} \right ] = - \int_{r_+}^\infty
\frac{f_0 Y^{\prime2}}{r^3Y^2} \leq 0
\label{anlwbd}
\ee
is negative.
Note that negativity of this integral does not imply existence of
a solution to the linearized equation near $T_c$, it is simply a necessary
condition. Since this integral is always negative at large
$T$, and positive as $T\to0$ (for $\kappa^2 \lesssim 0.4$), 
observing where it changes sign provides a lower bound on $T_c$. 
This bound gives an extremely reliable indicator of $T_c$ for a good
range of $\kappa^2$, well within the numerical range
we were able to explore.

The analytic bounds for $T_c$ have been plotted in figure \ref{fig:anbound}
together with the exact values of $T_c$ obtained by numerical computation.
The upper bound has only been shown for $\kappa^2 \leq 0.05$, as above this
value it becomes less predictive and clutters the plot, and indeed
beyond $\kappa^2\sim 0.2$ (corresponding to $q \sim 2.25$ 
in the notation of \cite{HHH}) it ceases to have quantitative
value for any $\alpha$. The lower bound
on the other hand becomes successively more accurate as the values of
$\kappa^2$ are stepped up, and gives a very good quantitative guide 
to the behaviour of $T_c$ as we vary $\alpha$ and $\kappa^2$.

It is easy to see that the effect of backreaction is to
decrease $T_c$ and thus make condensation harder.  
We can see this from the point of view of the bounds, as backreaction
at fixed $Q$ drops the temperature of the black hole and also
gives a smoother profile for $f_0$, which makes it harder for the 
bound (\ref{anlwbd}) to be satisfied.
Essentially, the effect of backreaction is that the condensation of the scalar
field not only screens the charge of the black hole, but also its mass,
as the scalar and gauge fields now contribute to the ADM mass. This means
that for a given charge and temperature, the radius of the black hole
is increased, which makes it harder for the scalar to condense.

One very interesting feature clearly exhibited in
the bounds is the turning point in $T_c$ as a function of $\alpha$ for
fixed $\kappa^2$. In the probe case, this occurs very near the Chern-Simons
limit $\alpha = L^2/4$, and is barely perceptible in the numerical data, 
however this is shown clearly in the upper bound in particular.
(We checked the numerical data by taking very small steps in the
$\alpha$ parameter near $L^2/4$, and were able to confirm this feature
in the numerical data, however, these points are not shown for clarity.)
Once backreaction is switched on, the minimum becomes much more pronounced,
and indeed for large backreaction ($\kappa^2 = 0.2$) the Chern-Simons
limit is showing a considerable enhancement of $T_c$ over the typical 
values for lower $\alpha$.

\FIGURE{
\includegraphics[width=10cm]{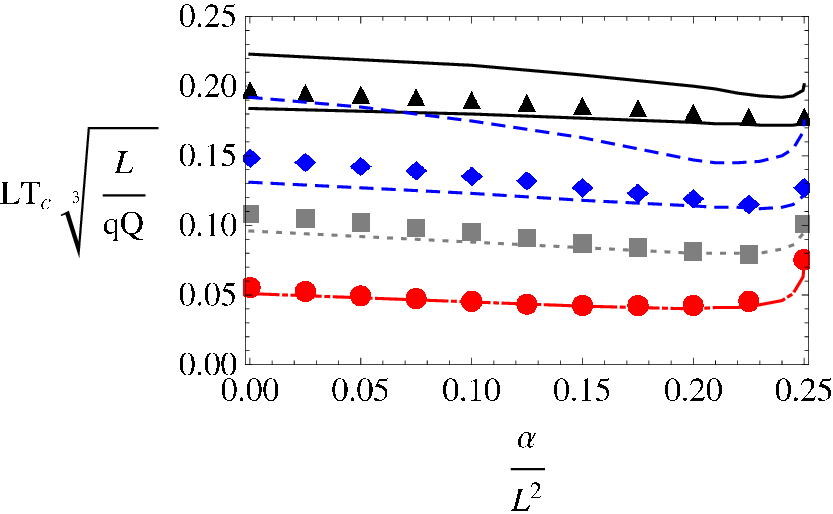}
\caption{
A plot of the critical temperature as a function of $\alpha$ for a
selection of $\kappa^2$. The analytic bounds are shown as lines and
the numerical data as points. Respectively: $\kappa^2=0$ is shown in
black, with solid lines and triangular data points; $\kappa^2 = 0.05$
has blue dashed lines and diamonds; $\kappa^2=0.1$ has a grey dotted 
line and squares; $\kappa^2=0.2$ has a red dot-dash line with circular 
data points. The lower bound is shown for all $\kappa^2$ values, but 
the upper bound is shown only for the lowest two values of $\kappa^2$, 
as they overlap significantly with the other data and confuse the plot.
}
\label{fig:anbound}
}

In order to find the actual behaviour of the bulk superconductor,
we integrated (\ref{phieq}$-$\ref{feq}) numerically. As already
stated, we took $L=Q=q=1$, and varied $r_+$ to study how the 
system reacted to varying temperature.
After solving for the matter and gravitational fields,
the temperature is computed and the
value of $\langle {\cal O}_3 \rangle$ read off.

Figure \ref{fig:TcPlot} shows $\langle {\cal O}_3 \rangle^{1/3}$ as a 
function of temperature for a variety of values of $\alpha$ and $\kappa^2$.  
Each line in the plot forms the characteristic curve 
of $\langle {\cal O}_3 \rangle^{1/3}$ condensing at 
some critical temperature. For simplicity we chose three values
of $\alpha$ to display the features of the system: the
Einstein limit ($\alpha=0$), the Chern-Simons limit ($\alpha=0.25$),
and the mid-point value $\alpha = 0.125$ to represent a `generic'
value of $\alpha$. We also only present the detailed information
for two values of $\kappa^2$: the case of no backreaction ($\kappa^2=0$),
and a backreaction of $\kappa^2 = 0.1$. The plots are qualitatively 
similar for other values of $\kappa^2$ and $\alpha$, and can be 
deduced from the information in figure \ref{fig:anbound} on critical
temperature.

Figure \ref{fig:TcPlot}(a), which displays the raw data, 
shows how varying $\alpha$ and $\kappa^2$ effects the 
height, shape and critical temperature of the condensate. This gives 
an alternate visualisation to figure \ref{fig:anbound} of the effect of both
backreaction and higher curvature terms on the critical temperature.
Note how the backreacting case clearly exhibits the special
nature of the Chern-Simons limit. Note also the disparity in the
unnormalized condensate value.  As in figure \ref{fig:anbound}, we 
see that increasing $\kappa^2$ reduces the 
critical temperature of the system markedly. We also see the 
effect of $\alpha$, which initially lowers $T_c$, then increases
it again for $\alpha$ approaching the Chern-Simons limit. Because we
are only showing the data for three values of $\alpha$, the
non-backreacting data do not display this effect, but it is clearly
present in the backreacting data, and indeed
increasing $\kappa^2$ enhances this effect, 

Figure \ref{fig:TcPlot}(b) shows the curves normalized by $T_c$. 
The effect of $\kappa^2$ is to increase the height of these graphs,
in spite of the fact that the raw data tends to have lower values
of $\langle {\cal O}_3 \rangle$. This is clearly because the most
significant impact of increasing gravitational backreaction is
that the critical temperature of the system is lowered.
In this case the effect of backreaction is
extremely marked, with the condensate varying more widely with $\alpha$.

\FIGURE{
\begin{tabular}{cc}
\includegraphics[width=5.2cm, angle=270]{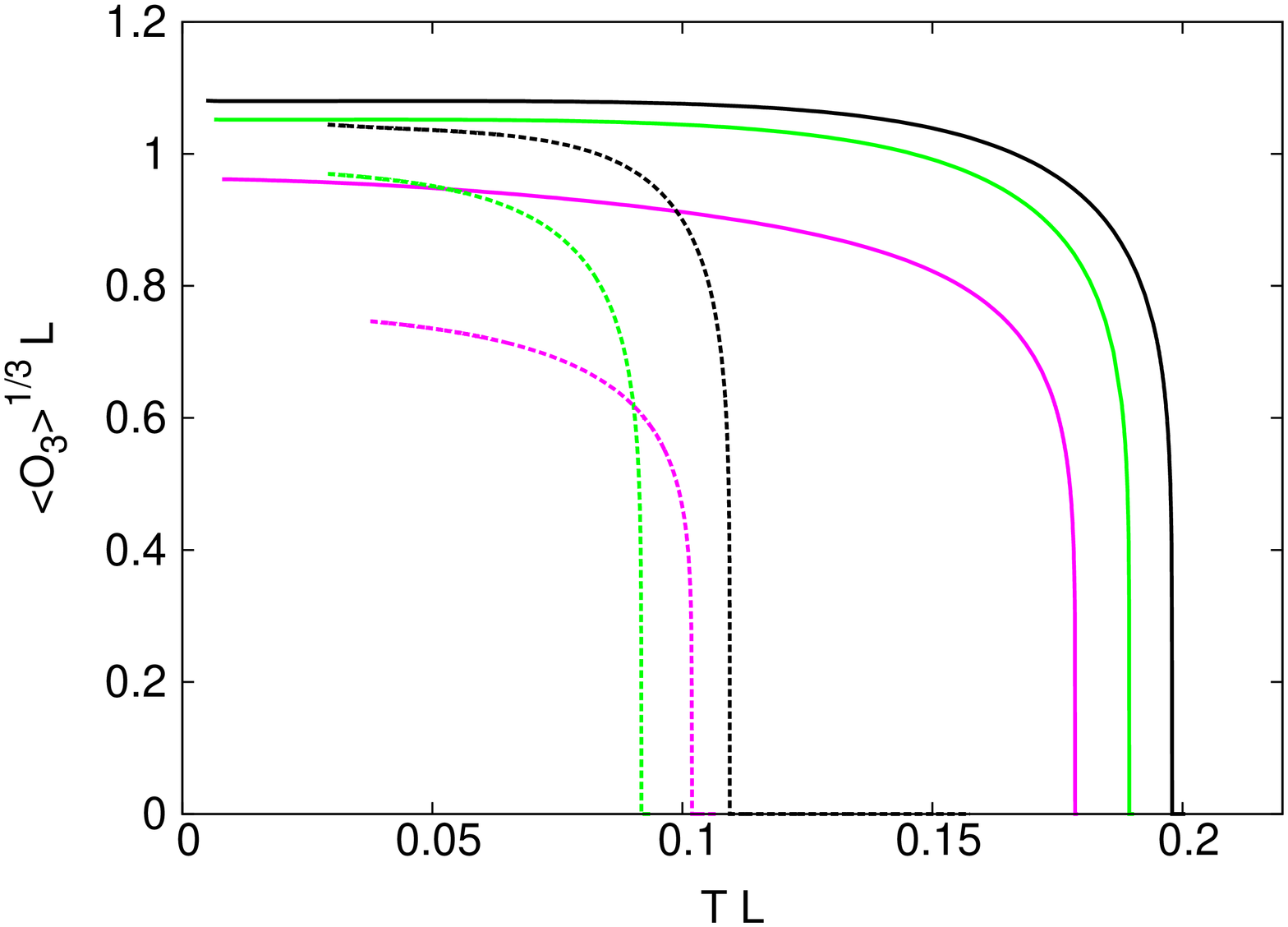} & 
\includegraphics[width=5.2cm, angle=270]{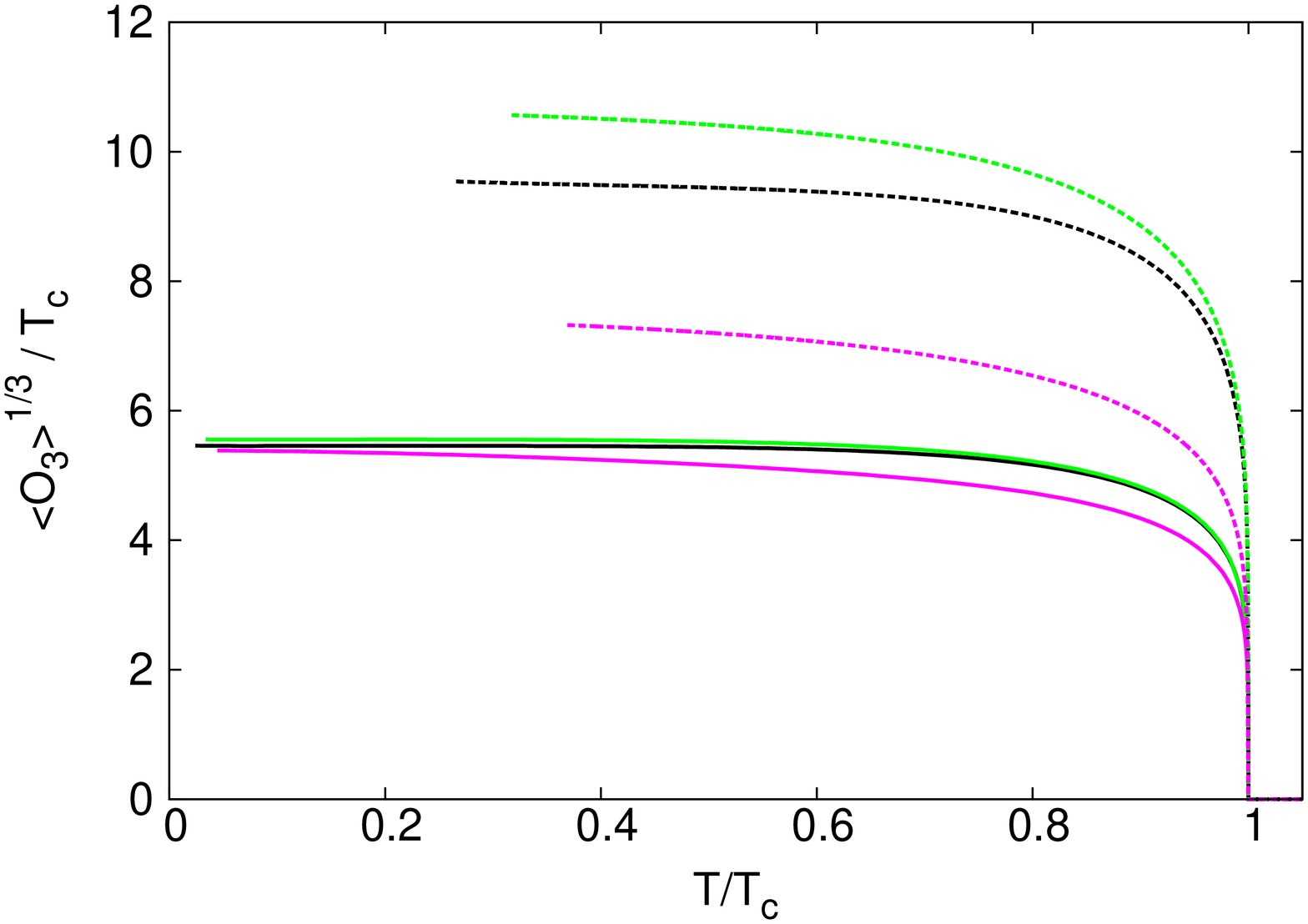} \\
(a) & (b)
\end{tabular}
\caption{Two plots of the condensate as a function of temperature for
a selection of values of $\alpha$ and $\kappa^2$. In each case, solid
lines correspond to $\kappa^2=0$ and dotted lines to $\kappa^2 = 0.1$.
The black plot is $\alpha=0$, green is $\alpha=0.125$ and magenta
is $\alpha=0.25$. The first plot shows unnormalized data, which indicates
the variation of critical temperature as both $\alpha$ and $\kappa^2$
vary.  Plot (b) shows the conventional plot of condensate against 
temperature, both rendered dimensionless by normalizing to $T_c$. 
}
\label{fig:TcPlot}
}

\section{Conductivity}
\label{sec:cond}

In \cite{HorRob}, Horowitz and Roberts observed an interesting 
phenomenon for the conductivity of the boundary theory.  They 
considered the model (\ref{action}) in the probe Einstein limit for a range
of different bulk scalar masses, and on computing the conductivity 
found an apparent universal relation
\begin{eqnarray}
\frac{\omega_g}{T_c} \simeq 8 \ ,
\end{eqnarray}
with deviations of less than 8 \%. In \cite{GKS} we
found evidence that this ``universality'' was not in fact stable
to the presence of stringy corrections, and we now wish to test 
the robustness of this result to backreaction.

Conductivity is conventionally expressed as the current density
response to an applied electric field:
\be
\sigma = \frac{\cal J}{\cal E} \, .
\label{condclass}
\ee
As the bulk field $A_\mu$ corresponds to a boundary four-current $J_\mu$ 
we must examine perturbations of $A_\mu$ to compute the conductivity.
Since we are dealing with full gravitational backreaction, we must
also perturb the metric, and compute the variation of the EGB gravity
equations. After some algebra we find:
\begin{eqnarray}
{\dot h}^\prime_{ti}-\frac{2}{r}{\dot h}_{ti}-{\ddot h}_{ri}
+\frac{L^2fe^{2\nu}}{r^2 - 2\alpha f}
\left( 1 - \frac{\alpha(2\nu'f + f')}{r} 
\right) \Delta h_{ri}
+\frac{2\kappa^2r^2 {\dot A}_i\phi^\prime}{r^2-2\alpha f} 
&=&0 \;\;\;
\label{ri}\\
\frac{e^{-\nu}}{rf} \left [ r f e^\nu A_i' \right ]'
-\frac{{\ddot A}_i}{f^2e^{2\nu}}
+\frac{L^2}{r^2f}\Delta A_i
-\frac{2}{f}q^2\psi^2A_i
+\frac{\phi'}{fe^{2\nu}}
\left( h_{ti}^\prime-\frac{2}{r}h_{ti} - {\dot h}_{ri} \right)
&=&0 \label{A1}
\end{eqnarray}
where $h_{ab}$ is the perturbation of the metric tensor, and
$A_i$ is the perturbation of the gauge field, which has only
spatial components.
Writing $A_i(t,r,x^i)=A(r)e^{i{\bf k}\cdot{\bf x}-i\omega t}e_i$, 
and setting ${\bf k}={\bf 0}$, we can integrate (\ref{ri}), and substitute
in (\ref{A1}) to obtain:
\be
A^{\prime\prime}+\left(\frac{f^\prime}{f}+\nu^\prime+\frac{1}{r}\right)A^\prime
+\left[\frac{\omega^2}{f^2e^{2\nu}}-\frac{2}{f}q^2\psi^2
-\frac{2\kappa^2r^2\phi^{\prime2}}{fe^{2\nu}\left(r^2-2\alpha f\right)}
\right]A =0 \;.
\label{A:eq}
\ee

We solve this under the physically imposed boundary condition of no
outgoing radiation at the horizon:
\be
A(r) \sim f(r)^{-i\frac{\omega}{4\pi T_+}} \ .
\label{ingoing}
\ee
Here, $T_+$ is the Hawking temperature defined at $r=r_+$.
In the asymptotic adS region $(r\rightarrow\infty)$, 
the general solution takes the form
\be 
\label{genasoln}
A=a_0 + \frac{a_2}{r^2}
+\frac{a_0 L_e^4 \omega^2}{2r^2}
\log \frac{r}{L}
\ee
where $a_0$ and $a_2$ are integration constants. 
Note there is an arbitrariness of scale in the logarithmic term,
as pointed out in \cite{HorRob}, however, this is related
to an arbitrariness in the holographic renormalization process
(see Appendix \ref{AppA})
and we present here the expression used in our
numerical computations to extract the behaviour of the gauge
field. 

To calculate the conductivity, we must therefore compute the current
dual to the gauge field (\ref{genasoln}), and its linear response
to an applied electric field. Since the details of the correspondence
are $\alpha$ dependent (the asymptotic adS lengthscale changes with
$\alpha$), we go through the explicit calculation in an appendix.
The result of this computation is\footnote{Note that the equation
for conductivity in \cite{GKS} was missing some factors of $L_e$:
(\ref{genasoln}) and (\ref{ConductivityEquation}) correct equations (4.4)
and (4.7) there.}
\be\label{ConductivityEquation}
\sigma=\frac{2a_2}{i\omega L_e^4 a_0 } 
+\frac{i\omega}{2} - i\omega \log \left ( \frac{L_e}{L} \right) \ .
\ee
Note that the imaginary term linear in $\omega$ has an arbitrariness
of scale from the counterterm subtraction, and in practise when we
present the plots of $\omega$ we will make use of this fact to choose
an appropriate renormalization scale to give the greatest transparency of the 
features inherent in the conductivity.

Figures \ref{fig:Conduct} and \ref{fig:omvTc} present various aspects
of our results for the conductivity at a range of values of $\alpha$ 
and $\kappa^2$. In figure \ref{fig:Conduct} we show the real and 
imaginary parts of the conductivity as a function of $\omega/T_c$
for no backreaction and a backreaction of $\kappa^2 = 0.05$
for our three sample values of $\alpha$: $0$, $0.125$, and $0.25$.

In each of these plots, the gap is clearly indicated by a rise in the 
real part of $\sigma$, which coincides with
the global minimum of $\text{Im}(\sigma)$.  As was noted in 
the derivation of (\ref{ConductivityEquation}), the imaginary part 
is only valid up to a linear term in $\omega$, the size of which is 
dependent on the renormalization scheme employed (and also on the 
charge $Q$).  Certainly, within the parameter range tested, one can 
always find an appropriate term to create a finite global minimum 
in $\text{Im}(\sigma)$ if it is not initially present. 
As in \cite{HorRob}, we suggest that the value of this minimum be 
taken as $\omega_g$, the value of the frequency gap. This has a clear
advantage over trying to determine the midpoint of the step in $\text{Re}
(\sigma)$, in that it gives an unambiguous value of $\omega_g$.
As the plots clearly show, once backreaction is included, the `step' 
becomes much more gentle and extended, and while the dip in 
$\text{Im}(\sigma)$ is smoothed, it is still clearly apparent, even
for the most extreme, backreacting Chern-Simons, case.
\FIGURE{
\includegraphics[width=5.2cm, angle=270]{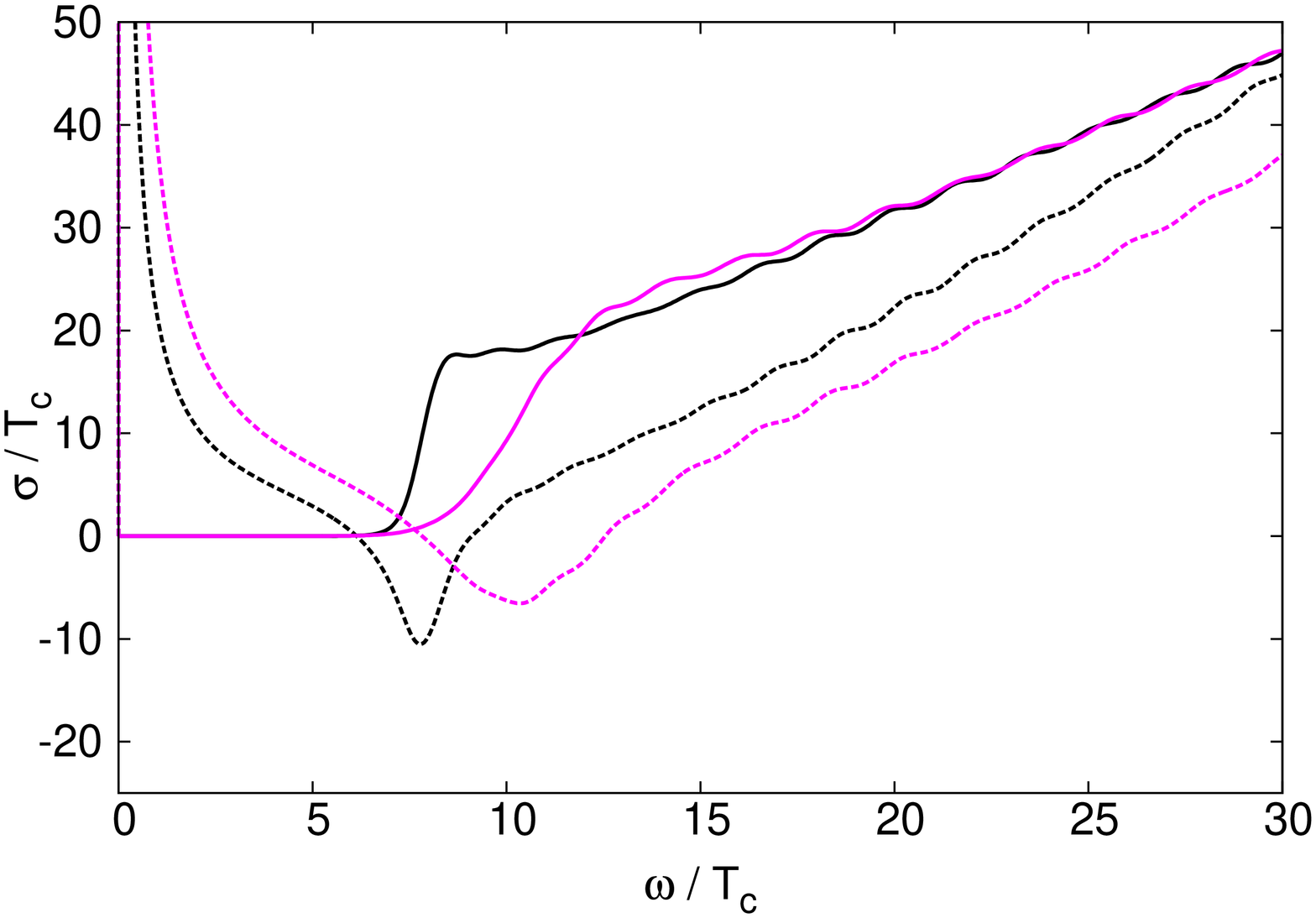}\\(a)
\begin{tabular}{cc}
\includegraphics[width=5.2cm, angle=270]{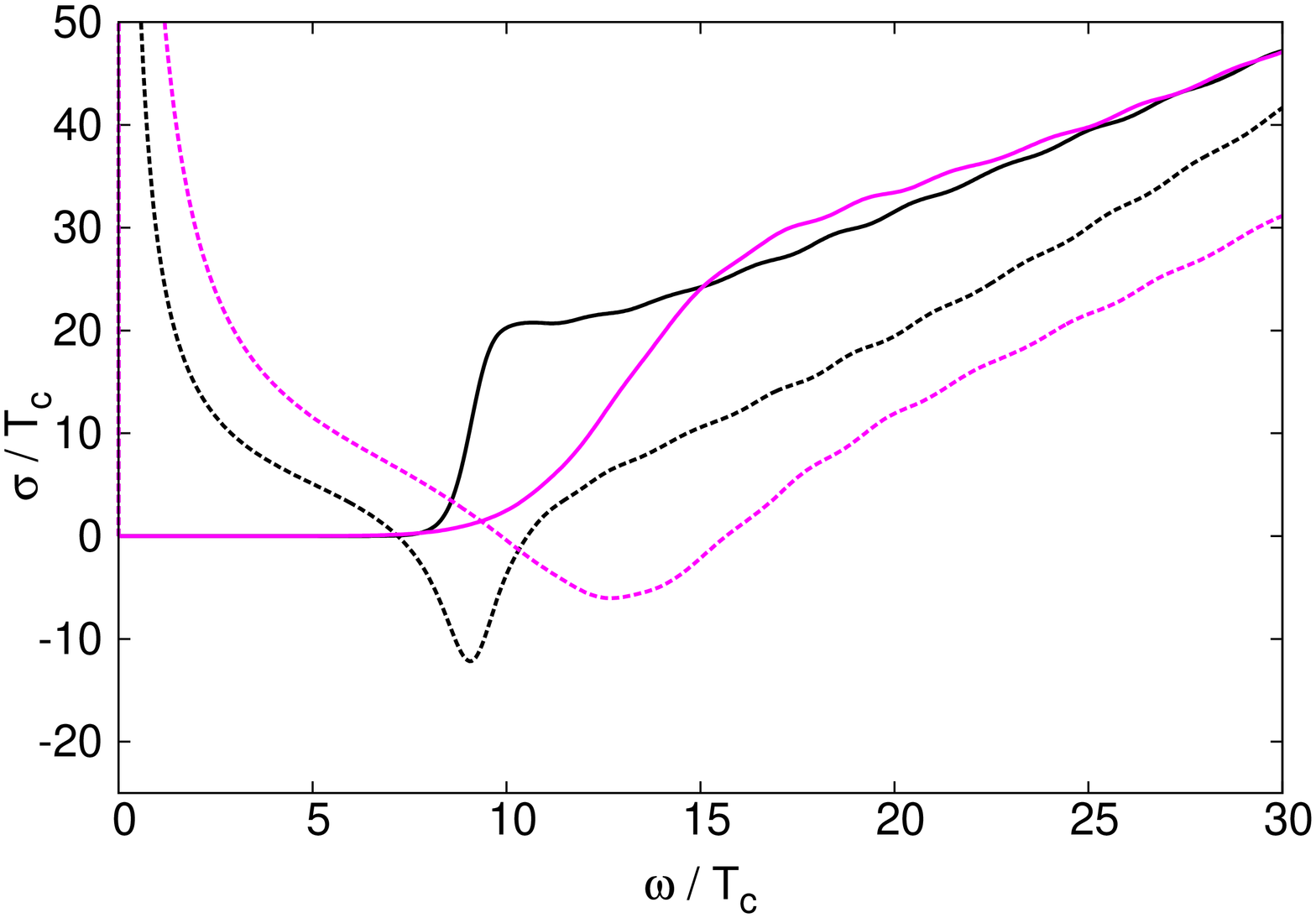} & 
\includegraphics[width=5.2cm, angle=270]{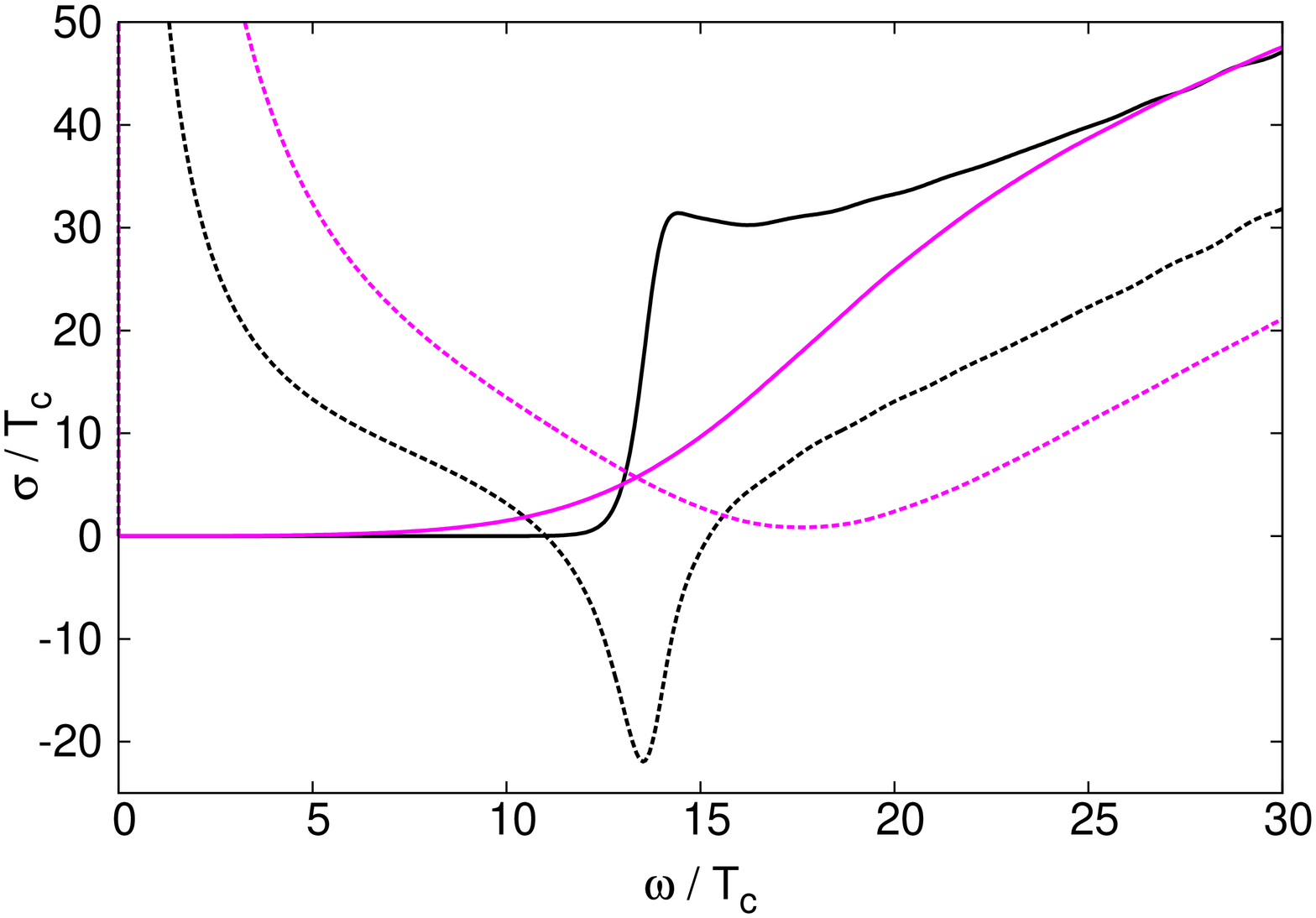} \\
(b) & (c)
\end{tabular}
\caption{Conductivity: a range of plots showing the real (solid line)
and imaginary (dashed line) parts of the conductivity as a function
of frequency. Both variables are normalized by $T_c$ to dimensionless 
parameters. In each case the conductivity is shown for no backreaction
in black, and for a backreacting parameter $\kappa^2 = 0.05$ in magenta.
Each plot represents a different representative value of $\alpha$:
(a) is $\alpha = 0$, (b) is $\alpha = 0.125$, and (c) is $\alpha=0.25$.
The slight undulations in the plots at large $\omega$ is a numerical
artefact due to the sensitivity of the system near the horizon.
}
\label{fig:Conduct}
}

The frequency gap is a distinct characteristic of a superconductor and in   
the BCS theory of superconductivity this frequency gap corresponds to the 
minimum energy required to break a Cooper pair.  As mentioned above, 
in \cite{HorRob} it was claimed that for the holographic superconductor 
the relation $\omega_g/T_c\simeq8$ had a certain universality, proving 
stable for a range of scalar masses and dimensions.  In \cite{GKS} this 
relation was shown to be unstable to Gauss-Bonnet corrections in the 
probe limit (for $m^2=-3/L^2$), and we unequivocally confirm
this feature. Figure \ref{fig:omvTc} gives a very clear indicator 
of how backreaction and higher curvature terms affect the gap.
Increasing either $\alpha$ or $\kappa^2$ increases $\omega_g/T_c$.
For the case of increasing $\alpha$, the effect occurs mainly because
of a shift in the gap, rather than a significant alteration
of $T_c$, which varies much more strongly with backreaction than
$\alpha$. On the other hand, varying $\kappa^2$ practically does not
alter $\omega_g$ at all, whereas $T_c$ drops dramatically, leading to 
a sharp rise in $\omega_g/T_c$.
\FIGURE{
\includegraphics[width=10cm]{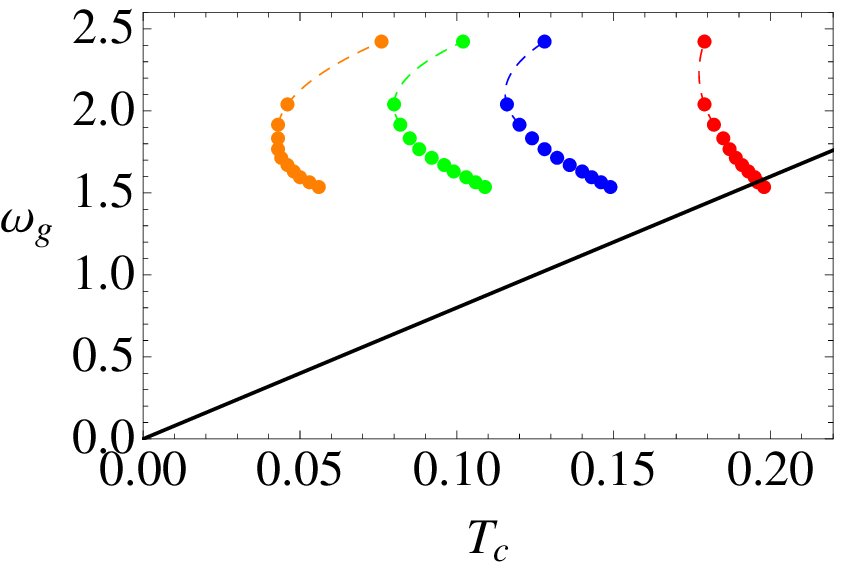}
\caption{The gap frequency as a function of $T_c$ (the line
$\omega_g=8T_c$ is shown in black). Each data point on the graph
represents a single pair $(\alpha,\kappa^2)$. The different colours 
represent different degrees of backreaction; from right to left:
Red is $\kappa^2 = 0$, Blue is $\kappa^2 = 0.05$, Green $0.1$, and
Orange $\kappa^2 = 0.2$. In each case $\alpha$ is incremented
from $0$ to $0.25$. As the gap alters rapidly near the Chern-Simons
limit, the dotted lines are added by hand to guide the eye.
}
\label{fig:omvTc}
}

\section{Conclusion}
\label{sec:disc}

The aim of this paper was to understand how including 
both Gauss-Bonnet corrections and backreaction might affect 
the holographic superconductor.  
Our results are clear: increasing backreaction
lowers the critical temperature of the superconductor hence 
increasing $\omega_g/T_c$. The effect of higher curvature terms
is more subtle. Although these initially act in a similar fashion
to backreaction in lowering the critical temperature, for
significant GB coupling the critical temperature eventually begins
to increase. The conductivity gap is also modified, with both 
$\omega_g$ and $T_c$ altering to increase the ratio $\omega_g/T_c$.
We have therefore unambiguously refuted the claim that there
is a universal gap $\omega_g \simeq 8T_c$ for these holographic
superconductors, as even in the Einstein limit there is no such
relation.

Our results show that there is a rich structure in higher dimensional
holographic superconductors, with or without higher curvature 
corrections. For simplicity, we focussed on a single value of 
the scalar mass which fixed the dimension of the boundary operator.
We believe this will clearly differentiate the effects of the backreaction
and higher curvature coupling. From the results of \cite{GBscs} we
do not expect any qualitative differences to appear from a varying
mass, although there will undoubtedly be quantitative differences,
particularly since the expressions for conductivity used are inaccurate
(as can easily be seen from dimensional grounds). It would be useful
to explore more fully this parameter space, and to see if altering the bulk 
potential further enhances the features of the system, as well as
exploring nonabelian gauge fields and more complex superconductors.

\acknowledgments
We would like to thank Jiro Soda for previous collaboration and
helpful input during this work. We would also like to thank Simon
Ross and Misao Sasaki for useful conversations. LB is supported by an STFC studentship, RG, SK and PMS acknowledge the support of STFC under the rolling grant ST/G000433/1.

\appendix
\section{Holographic renormalization}
\label{AppA}

In this, we follow the method of Skenderis \cite{Skenderis} and explicitly 
compute the counterterm and current dual to the electromagnetic bulk 
perturbation. First, we choose coordinates so that\footnote{Note that
the $h_{\mu\nu}$ here is distinct from the $h_{ab}$ metric perturbation
notation used in section \ref{sec:cond}!}
\be
ds^2 =  h_{\mu\nu}d\xi^\mu d\xi^\nu - L_e^2 \frac{d\rho^2}{\rho^2}
\ee
Clearly, asymptotically, $\rho = L_e^2/r^2 \to 0$, 
and $\xi^\mu = x^\mu/L_e$ with
\be
ds^2 \sim L_e^2 \left [ \frac{d\tau^2 - d\xi_i^2}{\rho} 
- \frac{d\rho^2}{\rho^2} \right ]
\ee
We then expand
\bea
h_{\mu\nu} &=& \frac{L_e^2}{\rho} \left [ \gamma^{(0)}_{\mu\nu} 
+ \rho^2 \gamma^{(4)}_{\mu\nu} \right ] \label{hexp}\\
A_\mu &=& L_e^{-1/2} \left [ A_\mu^{(0)} + \rho A_\mu^{(2)} 
+ \dots \right ] \label{Aexp}
\eea
and solve the equations of motion order by order in $\rho$.
(The factor of $L_e^{-1/2}$ ensures that the gauge fields
$A_\mu^{(n)}$ have the correct dimensionality.)

Focussing on the electromagnetic contribution, we can evaluate 
the action on-shell:
\bea
S &=& \int_{\cal M} - \frac{1}{4} F_{ab}^2 \sqrt{g} d^5x \nonumber\\
&=& \int_{\cal M} \frac{1}{2} A_b \nabla_a F^{ab} \sqrt{g} d^5x
- \int_{\partial{\cal M}} \frac{1}{2} F^{ab} n_a A_b \sqrt{h} d^4x \nonumber\\
&=& L_e \int_{\rho = \epsilon} A'_\mu A_\nu \gamma^{(0)\mu\nu} 
\sqrt{\gamma^{(0)}} d^4x
\eea
In these new coordinates, the gauge field equation of motion takes the form
\be
\rho\frac{d^2 A_i}{d\rho^2} - \frac{1}{4} \partial^2_{(0)} A_i= {\cal O}(\rho^2)
\ee
where $\partial^2_{(0)}$ represents the wave operator with respect to the
boundary metric $\gamma^{(0)}_{\mu\nu}$. The solution to this equation is
\be
A_i = L_e^{-1/2} A_i^{(0)} + \rho L_e^{-1/2}  \left [ A_i^{(2)} 
+ \frac{1}{4} \ln \rho\; \partial^2_{(0)}A_i^{(0)} \right ]
\label{Asol}
\ee
Thus the on-shell action,
\be
S = - \int_{\rho = \epsilon} A_i^{(0)} \left [ A_i^{(2)} 
+ \frac{1}{4} \partial^2_{(0)} A_i^{(0)}
+ \frac{1}{4} \ln \epsilon \; \partial^2_{(0)} A_i^{(0)} \right ] 
\sqrt{\gamma^{(0)}} d^4x \; ,
\ee
is logarithmically divergent. In order to find the correct counterterm,
we must invert the series solution (\ref{Asol}) to 
give $A^{(0)}_i = \sqrt{L_e}A_i + {\cal O}
(\epsilon)$, and hence obtain:
\bea
S_{ct} &=& \frac{L_e}{4}\ln \epsilon \int_{\rho = \epsilon} A_i \partial^2_{(0)} 
A_i \sqrt{\gamma^{(0)}} d^4x \nonumber \\
&=& \frac{L_e}{4}\ln \epsilon \int_{\rho = \epsilon} \frac{1}{2}
F_{\mu\nu}^2 \sqrt{h} d^4x
\eea

We can now compute the boundary current, which is given by 
the 1-point function
\be
\langle J^\mu \rangle = L_e^{-1/2} \frac{\delta S_{ren}}{\delta A_\mu}
\ee
Varying $S+S_{ct}$ explicitly gives
\bea
\delta S + \delta S_{ct} &=& \int_{\cal M} - F^{ab} \partial_a \delta A_b \sqrt{g} d^5x 
+ \frac{L_e}{2}\ln \epsilon \int_{\rho = \epsilon} F^{\mu\nu}
\partial_\mu \delta A_\nu \sqrt{h} d^4x \nonumber\\
&=& L_e \int_{\rho = \epsilon} \sqrt{\gamma^{(0)}} d^4x \delta A_i \left (
- 2 A_i'  + \frac{\ln \epsilon}{2} \; \partial^2_{(0)} A_i
\right )
\eea
Substituting for $A_i$ from (\ref{Asol}) gives
\bea
L_e^{-1/2} \frac{\delta S_{ren}}{\delta A_i} 
&=& - 2 \left [ A_i^{(2)} + \frac{1}{4} \ln \epsilon \; 
\partial_{(0)}^2 A_i^{(0)}
+ \frac{1}{4} \; \partial_{(0)}^2 A_i^{(0)} \right ]
+ \frac{\ln \epsilon}{2} \; \partial^2_{(0)} A_i^{(0)} \nonumber \\
&=&  -2 A_i^{(2)} - \frac{1}{2} \; \partial_{(0)}^2 A_i^{(0)} 
= -2 A_i^{(2)} + \frac{{\hat\omega}^2 - {\hat{\bf k}}^2}{2} \; A_i^{(0)} 
\label{adscurrent}
\eea
where the hatted quantities correspond to frequencies or wavenumbers
with respect to the dimensionless coordinates $\xi^i$.
Notice that a shift in the renormalization scale
$\epsilon \to \lambda \epsilon$,
results in a shift of the coefficient of the final term of $\ln\lambda/2$.

To get the conductivity, we can either compute the two-point function
following \cite{SonS}, or use the standard formula (\ref{condclass}),
noting that $E = {\dot A}_i^{(0)}$, to get the dimensionless conductivity
as
\be
{\hat \sigma} = \frac{2 A^{(2)}}{i\omega A^{(0)}}\bigg|_{{\bf k} = {\bf 0}} 
+\frac{i\omega}{2}
\ee

Finally, to obtain a dimensionally correct expression in terms of our 
original coordinates, we rewrite (\ref{Asol}) 
\bea
A_i &=& L_e^{-1/2} A_i^{(0)} + \frac{L_e^{3/2}}{r^2}  
\left [ A_i^{(2)} - \frac{1}{2} 
\ln \frac{r}{L_e} \partial^2_{(0)}A_i^{(0)} \right ] \label{Ar} \\
&=& L_e^{-1/2} \left [ A_i^{(0)} + \frac{L_e^{2}}{r^2}  \left ( A_i^{(2)}  
+ \ln (\frac{L_e}{L}) \frac{L_e^2({\bf k}^2 - \omega^2)}{2} A_i^{(0)} \right )
- \frac{L_e^4({\bf k}^2 - \omega^2)}{2r^2} \ln (\frac{r}{L}) A_i^{(0)} \right ]
\nonumber
\eea
from which we obtain
\be
\sigma = \frac{2 a_2}{i\omega L_e^4 a_0} 
- i\omega \ln (\frac{L_e}{L})
+\frac{i\omega}{2}
\ee
This is now the dimensionally correct expression for the conductivity,
although there is an ambiguity in the imaginary part of $i\omega\ln\lambda/2$
as already noted. This differs from the expressions in \cite{GKS,GBscs}
by factors of $L_e$, mainly due to an incorrect extrapolation of the
relation $G^R = -rfAA'|_{r\to\infty}$ from the Einstein limit.

\end{document}